\documentclass[preprint,prb]{revtex4}

\usepackage{graphicx,amsmath}
\usepackage{setspace}
\usepackage{enumerate}
\def\bsigma{\mbox{\boldmath $\sigma$}}
\def\bxi{\mbox{\boldmath $\xi$}}
\def\bomega{\mbox{\boldmath $\omega$}}
\def\bbomega{\mbox{\boldmath $\Omega$}}

\usepackage{amssymb}
\usepackage{amsmath}
\usepackage{graphicx}
\usepackage{dcolumn}
\usepackage{bm}
\newcommand{\bra}[1]{\langle#1|}
\newcommand{\ket}[1]{|#1\rangle}

\newcommand{\Tr}{{\rm Tr}}

\renewcommand{\d}{{\rm d}}

\mathchardef\mhyphen="2D 

\begin{document}

\title{Donor Spin Qubits in Ge-based Phononic Crystals}

\author{V. N. Smelyanskiy} 
\email{vadim.n.smelyanskiy@nasa.gov}
\affiliation{Quantum Artificial Intelligence Laboratory, NASA Ames Research Center, Mail Stop 269-3, Moffett Field, CA 94035.}
\author{V. V. Hafiychuk, F. T. Vasko}
\affiliation{SGT Inc., NASA Ames Research Center, Moffett Field, CA 94035.}
\author{A. G. Petukhov}
\affiliation{Physics Department, South Dakota School of Mines and Technology, Rapid City, SD 57701.}

\date{\today}

\begin{abstract}
We propose qubits based on shallow donor electron spins in germanium. Spin-orbit interaction for donor spins in germanium is in many orders of magnitude stronger than in
silicon. In a uniform bulk material it leads to very short spin lifetimes.
However the lifetime increases dramatically when the donor is placed
into a quasi-2D phononic crystal and the energy of the Zeeman
splitting is tuned to lie within a phonon bandgap. In this situation single
phonon processes are suppressed by energy conservation. The remaining
two-phonon decay channel is very slow. The Zeeman splitting within
the gap can be fine tuned to induce a strong, long-range coupling between the spins
of remote donors via exchange by virtual phonons. This, in turn, opens a very
efficient way to manipulate the quits.
We explore various geometries of phononic crystals in order to maximize the
coherent qubit-qubit coupling while keeping the decay rate
minimal. We find that phononic crystals with unit cell sizes of 100-150
nm are viable candidates for quantum computing applications and suggest several 
spin-resonance experiments to verify our theoretical predictions.
\end{abstract}
\maketitle

\section{Introduction}
Successful implementation of quantum information processing (QIP) requires not only invention of new quantum algorithms such as Shor algorithm, quantum error correction code or quantum adiabatic algorithm, but also further hardware development, i.e. realization of various qubit architectures - from trapped atoms to superconducting circuits.\cite{Ladd:2010} A significant advantage of solid state systems, based on different types of quantum dots\cite{Kloeffel:2013} or impurities in semiconductors,\cite{Kane:1998,Dutt:2007,Smelyanskiy:2005} is a capability to fabricate, manipulate and read out qubits using semiconductor nanotechnology and conventional electronics. On the other hand, all reliable and efficient QIP schemes simultaneously require both long qubit decoherence times\cite{Tyryshkin:2012} and controllable qubit manipulation, which poses a major challenge for practical implementation of these systems.

Indeed, shallow donor spin qubits in semiconductors have a number of advantages related to these requirements due to tunable spin-lattice interaction and a possibility to control spin states without a charge-induced noise. At the same time, broadly investigated silicon-based donor qubits with large spin dechorence times suffer from limitations in controlling and manipulating spins due to weak spin-orbit interaction.\cite{Claeys:2011}  Here we suggest a route for implementing new spin-qubit architectures based on donor spins embedded in specially crafted germanium structures (quasi-two-dimensional periodic phononic crystals or planar phonon waveguides) with large spin-orbit interaction of the Ge host and engineered phonon bangaps to simultaneously suppress spin decoherence and enable  strong spin-spin coupling between the qubits.

The large spin-orbit coupling inherent to shallow donors in bulk Ge enhances our ability to manipulate the spin qubits but could be detrimental for their coherence. In fact, the spin relaxation time of donors in Ge bulk is three to four orders of magnitude shorter than that in Si. To cope with this problem we will utilize Ge-based artificial periodic structures, known as {\em phononic crystals} (PHC).\cite{Liu:2000}  Similarly to the {\em photonic crystals}, that were invented to control the light,\cite{Yablonovich:1987} the phononic crystals of different dimensions can be used to control various types of acoustic waves, e.g. to filter and focus sound\cite{Yang:2004} or even to create the seismic proofing of buildings.\cite{Jia:2010} Recently, the state-of-the-art silicon-based phononic crystals have been fabricated\cite{Maldovan:2006,Alegre:2011,Safavi:2011} and several interesting physical effects such as the phonon-photon coupling  were experimentally demonstrated.\cite{Alegre:2011,Safavi:2011}

\begin{figure}[tb]
\begin{center}
\includegraphics[scale=0.75]{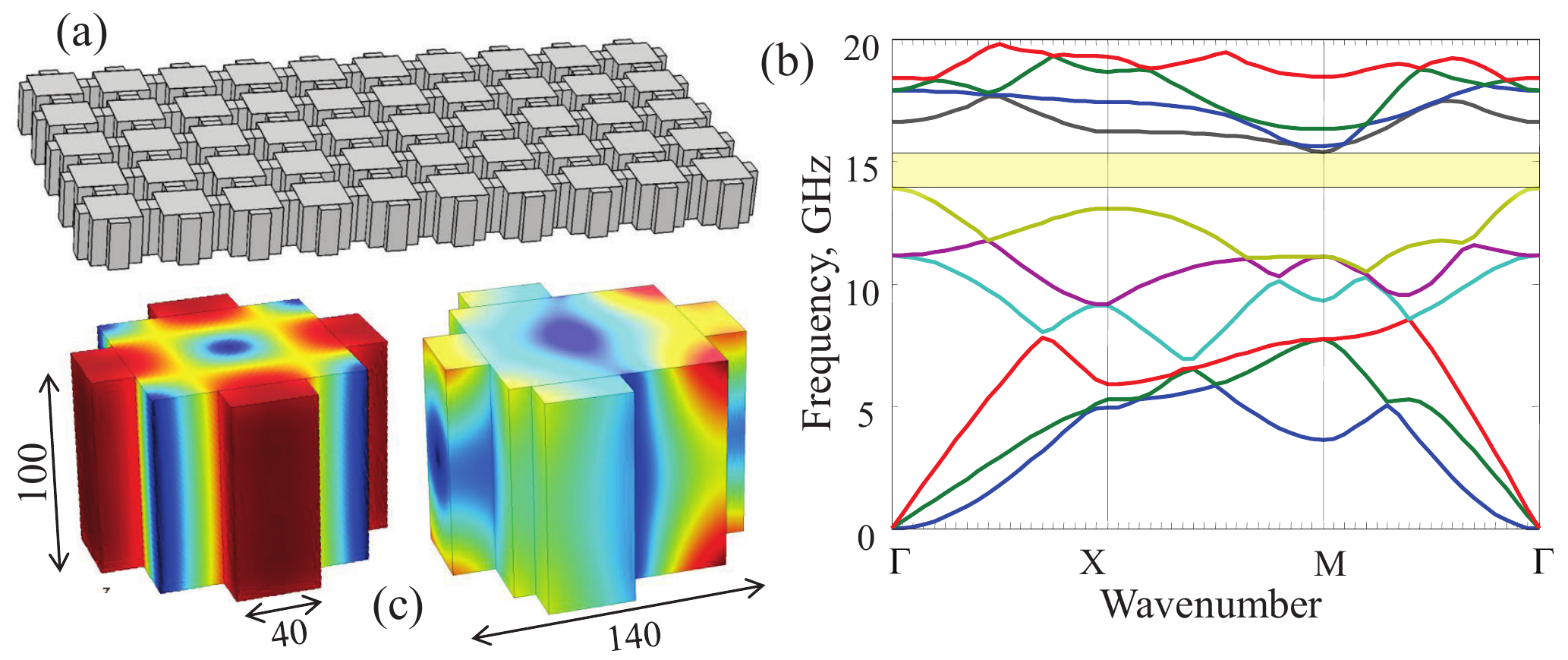}
\end{center}
\caption{
{\bf Phononic crystal.} ({\bf a}) Geometry of PHC formed by square lattice of holes. ({\bf b}) Dispersion laws along $\Gamma$X, $\Gamma$M, and XM directions with gap between 6th and 7th modes. ({\bf c}) Distributions of displacements over the unit cells of symmetric (left) and nonsymmetric (right PHCs for 7th mode at M-point (colored from blue, correspondent zero to red); sizes are in nm. }
\label{fig:PHC}
\end{figure}

\begin{figure}[thbp]
\begin{center}
\includegraphics[scale=0.7]{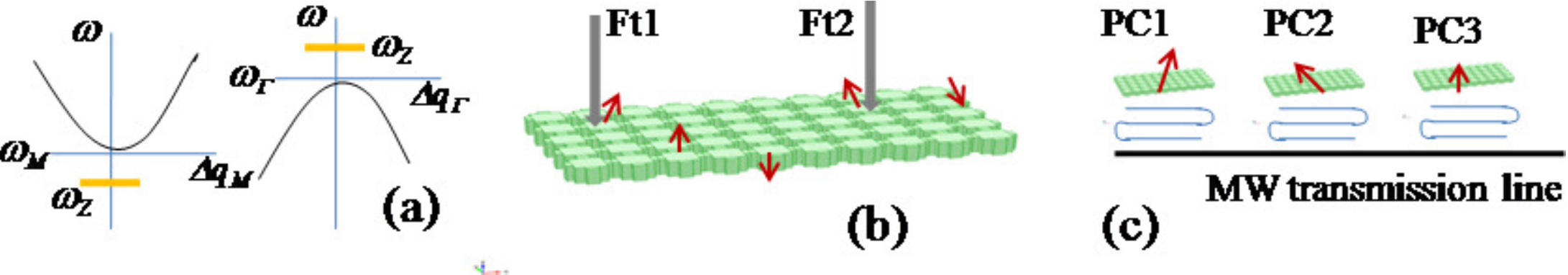}
\end{center}
\addvspace{-0.7 cm}
\caption{{\bf Qubit control.}({\bf a}) Dispersion curves around $M$ and $\Gamma$ points in parabolic approximation, where the REI-regime takes place if $\omega_Z\to \omega_{M,\Gamma}$. ({\bf b}) Manipulation of individual spins (red arrows) by inhomogeneous magnetic fields through ferromagnetic tips (Ft1 and Ft2). ({\bf c}) Schematic layout for manipulation of averaged spins $S_1$, $S_2$, $\ldots$ of PHC's register which are capacitively coupled to MW transmission line. }
\label{fig:qubit-control} 
\end{figure}

A proposed quasi-2D phononic crystal formed by a square lattice of cross-shaped holes in a suspended Ge layer is shown in Fig.~\ref{fig:PHC}a. This structure is similar to recently manufactured Si-based PHCs\cite{Alegre:2011,Safavi:2011} with $\sim$100 nm period and thickness defining the phonon gap within GHz frequency domain.  The phonon dispersion curves, shown in 
Fig.~\ref{fig:PHC}b, display pronounced gap in the frequency interval 13$\div$15 GHz.  If the Zeeman energy of the donor spin $\hbar\omega_Z$ is tuned inside the phonon gap the one-phonon spin-flip transitions will be forbidden due to energy conservation. As a result, the longitudinal relaxation rate, $\nu_1$, determined by very weak two-phonon processes, will be suppressed by five orders in magnitude compared to its bulk value. At the same time, the spin-lattice coupling will remain strong (3$\div$4 orders larger than in bulk silicon, depending on a donor position in the unit cell, Fig.~\ref{fig:PHC}c and the spin-spin interaction via virtual one-phonon exchange processes will exceed $\nu_1$ by many orders. If $\omega_Z$ is fine tuned and placed near the edge of the gap, see Fig.~\ref{fig:qubit-control}a,  both strength and lateral scale of a resonant exchange interaction (REI) enhance. Thus, a spin system with strong (and long-range for REI-regime) interaction is realized in PHC with suppressed relaxation (if a transverse rate $\nu_2$ remains only). Similar behavior is possible if Ge layer is sandwiched between rigid materials, when a quasi-gap appears due to weak penetration of vibrations between rigid and soft mediums.
\bigskip

The central result of this work is the description of the donor spin qubits with phonon mediated qubit-qubit interaction governed by the effective Hamiltonian given by Eq.~(\ref{eq:H}) below. 
The matrix $J^\bot_{nm}$ in Eq.~(\ref{eq:H}) is defined by the exchange of virtual phonons between $n_{th}$ and $m_{th}$ spins. The values of the matrix elements may exceed the two-phonon longitudinal spin decoherence rate $\nu_1$ by more than five orders in magnitude. This is possible because of the strong spin-orbit interaction and phonon gap engineering, which is tailored to completely eliminate the single-phonon decay. The fidelity factor of any quantum computing scheme is defined by the ratio $J/\nu_1 >10^{5}$ and this opens a way for a fault tolerant QIP. The main idea of our paper is that we can selectively turn the phonon-mediated interaction "on and off" by fine-tuning the Zeeman energies of individual qubits. The latter can be accomplished via local magnetic field sources such as a ferromagnetic AFM tip,\cite{Bode:2003} see Fig.~\ref{fig:qubit-control}b, because micrometer inter-donor scales in ultra-pure Ge\cite{Haller:1981} The pixel-like structure of PHC, which can be realized under a selective in-plane doping,\cite{Shinada:2005} provides a natural platform for the proposed quantum computing architecture. Another QIP scheme may be realized under connection of doped PHCs (with sizes $\geq 10~\mu$m and concentrations $\sim 10^{14}$ cm$^{-3}$) through the microwave transmitted line, when one can manipulate averaged spins ${\bf S}_1$, ${\bf S}_2$, $\ldots$, as shown in Fig.~\ref{fig:qubit-control}c (similar devices were demonstrated for SQUIDs\cite{VanLoo:2013} and for Si-based structures\cite{Sigillito:2014}). Below we will calculate the exchange matrix elements and propose a set of experiments to verify our theoretical predictions.

\begin{figure}[htbp]
\begin{center}
\includegraphics[scale=0.5]{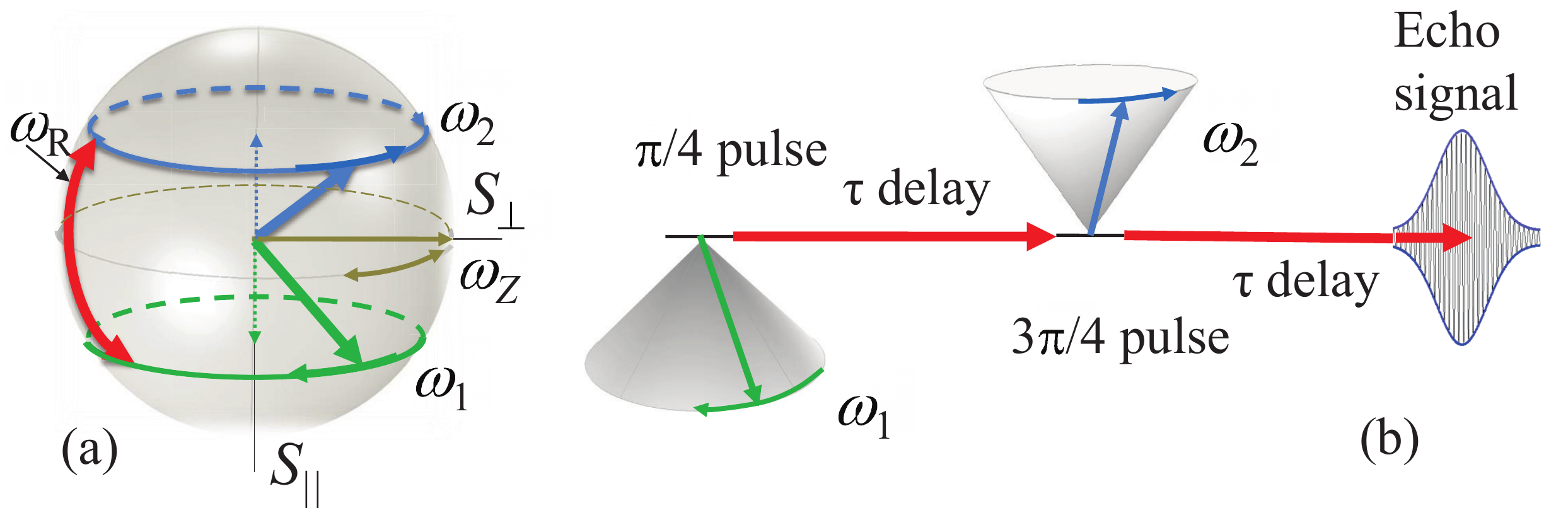}
\end{center}
\addvspace{-0.5 cm}
\caption{{\bf Spin dynamics under mw pumping.}  ({\bf a}) Bloch sphere for spin orientation ${\bf S}=({\bf S}_\bot,S_\|)$ with Zeeman frequencies renormalized due to exchange, $\omega_{1}<\omega_Z <\omega_{2}$, and Rabi oscillations frequency, $\omega_R$; here ${\bf S}_{t,\bot}^2 +S_{t,\|}^2 =$const due to the spin conservation law. ({\bf b}) Spin echo ($\pi /4 -\tau -3\pi /4 -\tau$) with $\tau$-dependent amplitude of echo signal due to difference of frequencies $\omega_{1}$ and $\omega_{2}$.  }
\label{fig:spin-dynamics}
\end{figure}

Both verification of the spin-Hamiltonian parameters and manipulation of spin coherence are possible under resonant microwave (mw) pumping of frequency $\omega\sim\omega_Z$. Under continuous mw pump, the exchange-renormalized $\omega_Z$ modifies a differential absorption shape in the linear response regime. An interplay between the Rabi oscillations frequency $\omega_R$ and the exchange contributions to Zeeman frequency ($\omega_{1,2}-\omega_Z$, see Fig.~\ref{fig:spin-dynamics}a) takes place for the nonlinear pumping case and it modifies a nonlinear differential absorption. If dephasing processes and long-range disorder are essential, a two-pulse spin echo measurements\cite{Schweiger:2001} enable to extract the exchange renormalization of $\omega_Z$. Fig.~\ref{fig:spin-dynamics}b shows the sequence ($\pi /4$ - $\tau$ - $3\pi /4$ - $\tau\to$ echo signal) with different frequencies of free rotation during $\tau$-delay intervals. As a result, an echo amplitude oscillates with $\tau$, if $\omega_{1,2}\neq \omega_Z$. Beside of this, a multi-pulse spin echo scheme can be applied for manipulation of averaged spin orientation.

\section{Results}
\subsection{Exchange Hamiltonian.} 
We consider a static magnetic field $\bm{B}=B\bm{h}$ ($\bm{h}$ is a unit vector) applied to
a Ge phononic crystal with donors occupying sites $\bm{r}_n$. The magnetic field induces Zeeman splitting $\hbar\omega_Z=g\mu_B B$ at each donor site. Here $\mu_B$ is the Bohr magneton and
$g=(g_\|+2 g_\bot)/3$ is electronic $g$-factor expressed through its longitudinal ($g_\|$) and transverse ($g_\bot$) principal components.\cite{Roth:1960,Hasegawa:1960}   
The Hamiltonian of the donor spin system, $\hat{H}=\hat H_Z+\hat H_{ph}+\hat{H}_{s\mhyphen ph}$,
comprising Zeeman, phonon,
and spin-phonon interaction terms, can be explicitly represented as: 
\begin{equation}
\label{eq:H}
\hat{H}=\frac{1}{2}\hbar\omega_Z
\sum_n\hat{{\bm\sigma}}_n{\bm h}
\!+\!
\sum_{{\bm q}\nu}\hbar\omega_{{\bm q}\nu}\left(b^\dagger_{{\bm q}\nu}b_{{\bm q}\nu}+\frac{1}{2}\right)
\!+\!\sum_{n,\alpha\ne \beta}\sum_{{\bm q}\nu}g_{\bm{q}\nu}
\epsilon_{n,\alpha\beta}^{{\bm q}\nu}\hat\sigma_{n,\alpha} h_\beta e^{i\bm{qr}_n}\left(b_{{\bm q}\nu}+b^\dagger_{-{\bm q}\nu}\right)\!,
\end{equation}
where $\hat\sigma_n^\alpha$ are Pauli matrices with $\alpha=x,y,z$ along Ge crystal axes [100], [010], and [001].  Bosonic creation (annihilation) operators 
$b_{\bm{q}\nu}^\dagger$ ($b_{\bm{q}\nu}$) correspond to a phonon mode $\bm{q}\nu$ with frequency $\omega_{\bm{q}\nu}$ and wave vector $\bm{q}$ 
within a two-dimensional Brillouin zone of the phononic crystal.
The spin-phonon coupling constant $g_{\bm{q}\nu}$ is defined as $g_{\bm{q}\nu}=\hbar\omega_Z K (1/2\varrho_M V\omega_{\bm{q}\nu})^{1/2}$, where $\varrho_M$ is an
 average mass density of the phononic crystal, $V$ is a
normalization volume, and $K=
2\left(g_\|-g_\bot \right)\Xi_u/9g\Delta_0$, where $\Xi_u$ and $\Delta_0$ are the shear deformation potential of bulk Ge and the valley-orbit splitting of a Ge donor, respectively.\cite{Roth:1960,Hasegawa:1960,Wilson:1964} The dimensionless coupling constant $K$ characterizes strength of the spin-phonon interaction relative to $\hbar\omega_Z$. Finally, the quantity 
\begin{equation}
\epsilon_{n,\alpha\beta}^{{\bm q}\nu}=i
\int |\psi_n(\bm{r})|^2e^{i\bm{q}(\bm{r}-\bm{r}_n)}
\left[\hat{\pi}_\alpha e_\beta^{\bm{q}\nu}(\bm{r})+\hat{\pi}_\beta e_\alpha^{\bm{q}\nu}(\bm{r})\right]d\bm{r}
\label{eq:strain}
\end{equation}
is the dynamic strain tensor of the phonon mode ${\bm q}\nu$ averaged with the fully symmetric 
wave function $\psi_n(\bm{r})=\psi_{A_1}(\bm{r}-\bm{r}_n)$ of the donor ground state $A_1$.\cite{Roth:1960,Hasegawa:1960} Here $\hat\pi_\alpha=q_\alpha-i\partial/\partial x_\alpha$, and the phonon polarization vector $e_\alpha^{\bm{q}\nu}(\bm{r})$ is the eigenvector of the elastic eigenvalue problem for the phononic crystal.

Hamiltonian (\ref{eq:H}) has several peculiarities that make donor arrays in Ge phononic crystals stand out compared with similar systems based on other materials such as Si.  It is worth mentioning that the coupling of the ground-state Zeeman doublet with the elastic strain is enabled by the $g$-factor
anisotropy $\Delta g=g_\bot-g_\|$, which, in turn, is induced by the spin-orbit interaction and depends on the effective
mass anisotropy. The latter two factors are much stronger in Ge
($\Delta g/g\simeq 0.5$) than in Si ($\Delta g/g\simeq 1.5\times 10^{-3}$). As a result, the coupling constant $K$ in Ge ranges from $1\times 10^3$ for
As donor to $7.5\times 10^3$ for Sb donor, exceeding similar values for Si by about three
orders in magnitude. Second, because of the valley symmetry, the spin-orbit interaction in Ge couples the donor spins only to shear components of the strain tensor $\varepsilon_{xy}$,   $\varepsilon_{xz}$, etc. (note that $\alpha\ne\beta$ in Eq.~(\ref{eq:H})).  This creates a simple way to protect the spin qubits from detrimental effects of static random strains.
Indeed, the random strain Hamiltonian is similar to the last term in Eq.~(\ref{eq:H}):  $H_r=\hbar K\omega_Z
\sum_{n,\alpha\ne \beta}
\varepsilon_{\alpha\beta}(\bm{r}_n)\hat\sigma_{n,\alpha} h_\beta$, where
$\varepsilon_{\alpha\beta}(\bm{r}_n)$ is a random quantity fluctuating from site to site. For the magnetic field in $z$-direction $H_r$ does
not contain $\hat\sigma_z$, which implies that the first order correction to $g$ vanishes,\cite{Roth:1960} since $\delta g \propto 
\bra{\pm} H_r \ket{\pm}=0$ (here $\ket{\pm}$ are eigenstates of $\hat\sigma_z$). For a typical random strain fluctuation  with r.m.s.  $\delta\varepsilon\sim 10^{-6}$,\cite{Wilson:1964} the second oder correction, $\delta g/g\sim (K\delta\varepsilon)^2\simeq 10^{-6}$ is negligible. This finding is confirmed by experimental observations revealing  
strong anisotropy in the inhomogeneous ESR line-broadening for Ge donors due to random strains~
and a drastic narrowing of the lines for [001] direction of the magnetic field.\cite{Wilson:1964} 

Based on the above we will assume that the magnetic field is perpendicular to the plane of the phononic crystal (to eliminate random strain noise) and that $\omega_Z$ lies inside the phonon band gap (to suppress single-phonon decoherence processes). In what follows we will use the interaction picture with unperturbed Hamiltonian $\hat H_0=
\hat{H}_Z +\hat{H}_{ph}$ and $\hat H_{s\mhyphen ph}$ taken as a perturbation.
The effective qubit-qubit Hamiltonian can be derived using the Redfield equation\cite{Redfield:1957,Breuer:2007} for the reduced density matrix $\hat{\rho}_s=\mathrm{Tr}_{\hat{H}_{ph}}\hat{\rho}$ and employing the rotating wave approximation (RWA): 
\begin{equation}
{\d\hat{\rho}_s(t)\over\d t}=-{1\over\hbar^2}\int_0^\infty\d \tau\,\Tr_{\hat{H}_{ph}}
[\hat{H}_{s\mhyphen ph}(t),[\hat{H}_{s \mhyphen ph}(t-\tau),\hat{\rho}_s(t)\otimes\hat{\rho}_{ph}]]
\simeq \frac{i}{\hbar}\left[\hat{\rho}_s(t),\hat{H}_{LS}+\hat{H}_{s\mhyphen s}\right],
\label{eq:Redfield}
\end{equation}
where
\begin{equation}
\label{eq:Hss}
H_{s\mhyphen s}=\sum_{n,m\ne n}J^\bot_{nm}\sigma_n^+\sigma_m^-
\end{equation}
with
\begin{equation}
\label{eq:Jnm}
J_{nm}^\bot=\frac{\hbar^2\omega_Z^2 K^2}{\varrho_M V}\sum_{{\bm q}\nu}\frac{\xi_{n}^{\bm{q}\nu}
\xi_{m}^{\bm{q}\nu*}}{\omega_Z^2-\omega_{\bm{q}\nu}^2}
e^{i\bm{q}(\bm{r}_n-\bm{r}_m)},
\end{equation}
and $\xi_{n}^{\bm{q}\nu}=\epsilon_{n,xz}^{\bm{q}\nu}-i\epsilon_{n,yz}^{\bm{q}\nu}$. 
In Eq.~(\ref{eq:Redfield}) the first equality is the initial Redfield equation and the second equality is the final result obtained after tracing over the phonon degrees of freedom and applying RWA, which means that the time average of all quickly oscillating processes is zero.  The Hamiltonian $H_{LS}$, which  is a complete analog of the Lamb shift Hamiltonian in quantum electrodynamics, has the same form as $H_{s\mhyphen s}$ but with $n=m$. 
The Hamiltonian
$H_{LS}$ can be absorbed into the unperturbed Hamiltonian Hamiltonian $\hat{H}_0$, which leads to a renormalization of $\omega_Z$ in $H_Z$ and in Eq.~(\ref{eq:Jnm}). 
Our numerical estimates show that this renormalization $\delta\omega_Z$ is insignificant and constitutes at most $\delta\omega_Z/\omega_Z \sim 10^{-4}$.  Thus we will concentrate on the
interqubit Hamiltonian $\hat{H}_{s\mhyphen s}$ and use renormalized values of $\omega_Z\rightarrow 
\omega_Z+\delta\omega_Z$ in Eq.~(\ref{eq:Jnm}). 
Notably, the dynamics of the system with $\omega_Z$ in the gap is non-dissipative to 
the second order in $g_{\bm{q}\nu}$ and is governed
by the exchange Hamiltonian $H_{s\mhyphen s}$ in agreement with our initial claim.

The Hamiltonian (\ref{eq:Hss}) lays out a foundation of the proposed QIP architecture.
Before proceedings with QIP applications of the Hamiltonian (\ref{eq:Hss})  we will address physical implications of 
Eqs~(\ref{eq:Hss}) and (\ref{eq:Jnm}), 
investigate asymptotic behavior  of the exchange integrals, and estimate spin decoherence time due to two-phonon processes.  The most interesting property of the phonon mediated exchange is a possibility to tune  magnitude and the range of qubit-qubit interaction by changing the Zeeman splitting.

\subsection{Properties of donor-based spin qubits.} 
To better understand the properties of the Hamiltonian (\ref{eq:Hss}) we will examine asymptotic behavior of the exchange integrals. Without any loss of generality we assume that near extrema of the phonon band $\bm{q}_0$ (either $\Gamma$ or $M$-points of the Brillouin 
zone) $\omega(\bm{q})\simeq\omega_0(1+l^2\delta q^2/2)$, where $\delta\bm{q}=\bm{q}-\bm{q}_0$ and  we introduced a characteristic length $l$, 
which describes dispersion of 
$\omega(\bm{q})$ near $\bm{q}_0$.  Since $\bm{q}$ is two-dimensional the exchange
integrals depend only on a projection $\bm{\rho}_{nm}$ of $\bm{r}_{nm}=\bm{r}_n-\bm{r}_m$ onto
the $(x,y)$-plane of the phononic crystal. It is convenient to introduce another ($\omega$-dependent) length parameter $r_0=l\omega_0/\sqrt{\omega_0^2-\omega_Z^2}$, defining the range of the qubit-qubit interaction.  Replacing summation in Eq.~(\ref{eq:Jnm}) with integration  and assuming
large interqubit separation, $\rho_{nm}\gg  r_0$, we can extend the upper integration limit to 
infinity and obtain:
\begin{equation}
J_{nm}^\bot \simeq J_0^\bot K_0\left(\frac{\rho_{nm}}{r_0}\right)\sim  J_0^\bot 
\sqrt{\frac{\pi r_0}{\rho_{nm}}}\exp\left(-\frac{\rho_{nm}}{r_0}\right),  \hskip 0.75 cm J_0^\bot=-\frac{\hbar^2\omega_Z^2 K^2\xi_0^2}{2\pi\omega_0^2\varrho_M l^2 d}.
\label{eq:large-dist}
\end{equation}
Here $K_0(x)$ is the modified Bessel function of the second kind, $d$ is the period of the PHC lattice, which also equals to its thickness, and we neglected weak  $\bm{q}$-dependence  of $\xi_{n\nu}^{\bm{q}}\simeq \xi_0\sim \pi/d$. The pre-factor $J_0^\bot/2\pi\hbar$ ranges from 100~KHz (As donors) 
to 1.7~MHz (Sb donors). 
\begin{figure}[htbp]
\begin{center}
\includegraphics[scale=0.35]{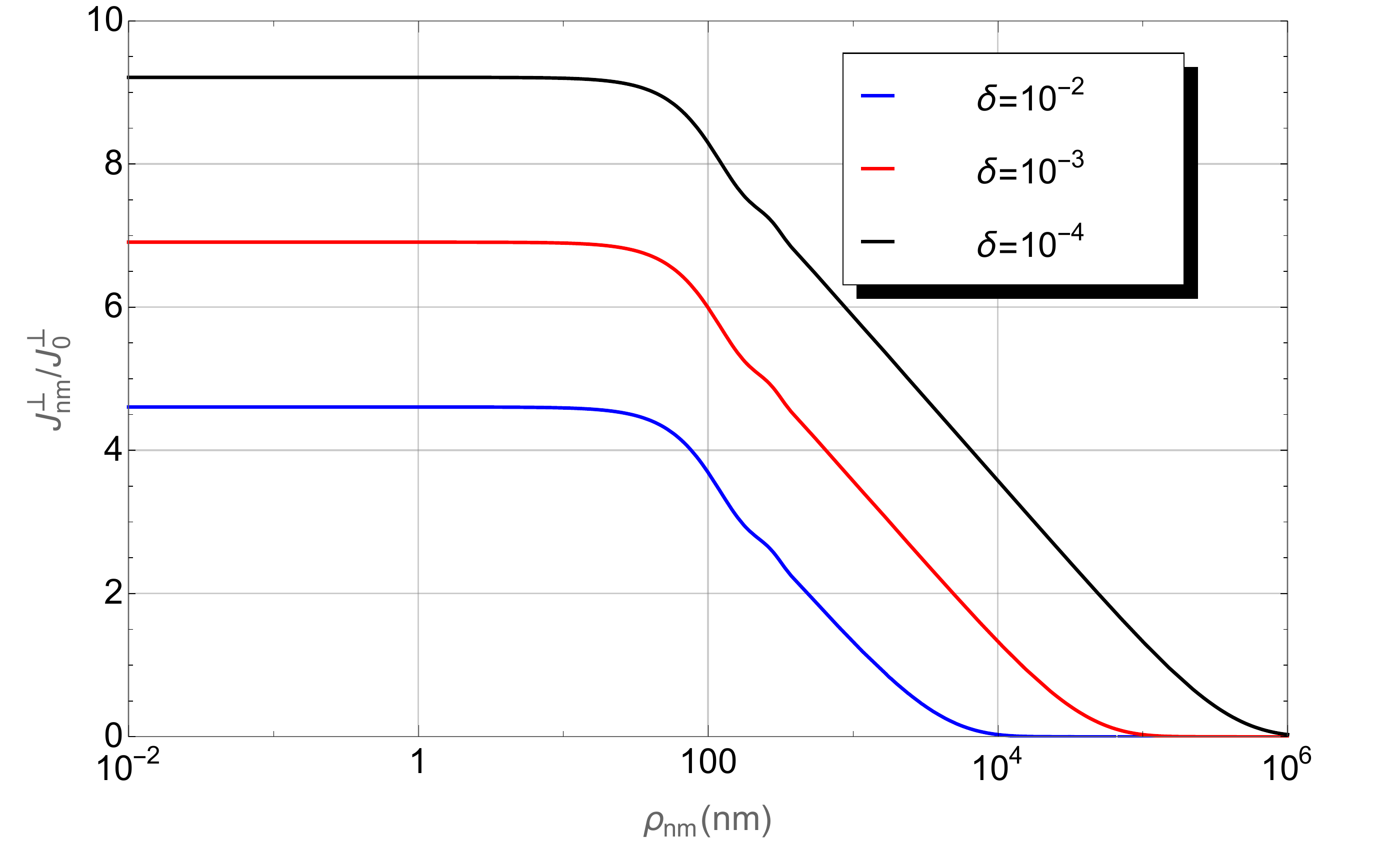}
\end{center}
\addvspace{-0.7 cm}
\caption{ {\bf Exchange integrals for different values of $\delta=\omega_0/\sqrt{\omega_0^2-
\omega_Z^2}$.}}
\label{fig:exchange}
\end{figure}
At intermediate distances $\rho_{nm}<r_0$ we have to cut-off the upper limit near the edge of the Brillouin zone
$q_c\sim \pi/d$. This yields
\begin{equation}
\label{eq:intermediate-dist}
J^\bot
_{nm}\simeq J_0^\bot\!\left[\log \sqrt{1 + q_c^2 r_0^2}-\frac{q_c^2 \rho_{nm}^2}{8}\cdot\,  _2F_3\left(1,1;2,2,2;-\frac{q_c^2 \rho_{nm}^2}{8}  \right)\!\right],
\end{equation}
where $_pF_q(a;b;x)$ is the generalized hypergeometric function. Eq.~(\ref{eq:intermediate-dist}) describes
$J^\bot_{nm}$ very accurately apart from the exponential tail at large $\rho_{nm}$ as in Eq.~(\ref{eq:large-dist}). If $1/q_c\ll \rho_{nm} < r_0$ the function in the right hand side of Eq.~(\ref{eq:intermediate-dist}) assumes a simple logarithmic form
$J^\bot_{nm}=
J_0^\bot\log(\rho_{nm}/r_0)$. The results of numerical calculations of $J_{nm}^\bot$ are shown
in Fig.~\ref{fig:exchange}. Strikingly, the qubit-qubit interaction is long-range due to a very weak logarithmic dependence of the exchange integrals on the interqubit separation for $\rho_{nm} < r_0$.
Since $r_0$ becomes very large one can execute a SWAP-gate\cite{Barenco:1995} between the qubits separated by the distances as large as 0.1 cm! As we mentioned before, the qubit-qubit interaction and, in turn, the execution of the SWAP gate are enabled by the resonant energy transfer 
when the double spin-flip process is accompanied by the simultaneous emission and absorption of a virtual phonon if the Zeeman energies of the two qubits are brought into resonance with each other. Furthermore, when these energies are near the edge of the phonon band gap the interaction becomes extremely long-range.     

Finally, we calculate the longitudinal relaxation rate $\nu_1$ 
at low ($\ll \hbar\omega_Z$) temperatures, which is determined by the two-phonon spin-flip transitions via emission of modes $(\nu,{\bf q})$ and $(\nu_1,{\bf q}_1)$. Using the golden rule with the forth-order ($\propto K^4$) matrix elements we obtain:
\begin{equation}
\nu_1 =\frac{2\pi}{\hbar}\sum\limits_{\nu {\bf q}\nu_1 {\bf q}_1}g_{\nu{\bf q}}^2 g_{\nu_1{\bf q}_1}^2\left| {\frac{\xi^{(\nu {\bf q})}\cdot\bxi^{(\nu_1{\bf q}_1 )}}{\hbar\omega_Z -\hbar\omega_{\nu {\bf q}} }}+(\nu{\bf q}\leftrightarrow \nu_1{\bf q}_1 )\right|^2 
\delta (\hbar\omega_Z -\hbar\omega_{\nu {\bf q}}-\hbar\omega_{\nu_1 {\bf q}_1}) .
\label{eq:nu1}
\end{equation}
Using Eq.~(\ref{eq:nu1}) we can estimate 
$\nu _1 /\omega _Z$ as
\begin{equation}
\frac{\nu_1}{\omega_Z}\sim \left(\frac{K^2 \hbar \omega _Z^3}{\varrho_M d\cdot v_s^4} \right)^2 ~~, 
\label{eq:nu-omega}
\end{equation}
where $v_s$ is an average speed of sound. For the above parameters $\nu_1\propto\omega_Z^7$ and longitudinal relaxation time $\nu_1^{-1}$ exceeds hundreds seconds. Thus, $J_{nm}/\hbar\nu_1$ exceeds $10^5$, and for $\omega_Z$ within the phononic gap, $\nu_1$ is negligible.


\subsection{Spin Dynamics and Bloch equation.} Temporal evolution of weakly interacted spins randomly placed into PHC is described by the averaged spin vectors,\cite{Dyakonov:2008} ${\bf s}_{kt}=(1/2){\rm Tr}_S \hat{\bsigma}_k \hat{\rho}_t$, where $\hat\rho_t$ is the multi-spin density matrix governed by the equation with the Hamiltonian $\hat H_{eff}$. We restrict ourselves by the second order accuracy on $\hat H_{SS}$ (the mean field approximation) and factorize the two-spin correlation function as ${\rm Tr}_S\hat\rho_t\hat\sigma_{k,\alpha}\hat\sigma_{k',\beta}\approx s_{kt,\alpha}s_{k't,\beta}$. 
As a result, the system of the nonlinear Bloch equations for ${\bf s}_{kt}$ takes form:
\begin{equation}
\frac{d{\bf s}_{kt}}{dt} +\hat\gamma\cdot{\bf s}_{kt}=\left[\left( \bomega_{Zk} +\Delta\bomega _{t} \right)  \times {\bf s}_{kt} \right]   
-\frac{2}{\hbar} \sum\limits_{k'(k'\neq k)}\left[\left(\hat J_{kk'}\cdot {\bf s}_{k't}\right)\times {\bf s}_{kt} \right] ~.
\label{eq:Bloch}
\end{equation}
Here $\Delta\bomega_t$ is the time-dependent Zeeman frequency under a mw pumping (below $\Delta\bomega_{t}\bot 0Z$), $\hat\gamma\cdot{\bf s}_{kt}$ describes relaxation of $k$th spin, and the last term gives the effective exchange contribution written through the matrix (\ref{eq:Jnm}). Under pumping $\Delta\bomega_{Zt}$ switched on at $t=0$ Eqs. (\ref{eq:Bloch}) should be solved with the initial conditions ${\bf s}_{kt=0}=(0,0,s_{k0})$ where $s_{k0}\simeq -1/2$ if temperature $\ll\hbar\omega_Z$ and $s_{k0}\to 0$ for the high temperature ($\gg\hbar\omega_{Zk}$) region.

Instead of microscopic set $\{ {\bf s}_{kt}\}$, we consider the spin orientation  ${\bf S}_{{\bm r}t}=\left\langle\sum\nolimits_k {\delta ({\bf r}-{\bf r}_k ){\bf s}_{kt}}\right\rangle /$ $\left\langle\sum\nolimits_k {\delta ({\bf r}-{\bf r}_k )}\right\rangle$ where $\left\langle  \cdots  \right\rangle  = V^{-N}\int {d{\bf r}_1  \ldots } \int {d{\bf r}_N  \cdots }$ stands for the averaging over $N$ donors in volume $V$. For the large ($\gg d,a$) scale inhomogeneity case, ${\bf S}_{{\bm r}t}$ is weakly dependent on transverse coordinate $z$ and is governed by the spin diffusion equation
\begin{align}
\frac{d{\bf S}_{{\bm r}t}}{dt} +\hat\gamma\cdot{\bf S}_{{\bm r}t}&= \left[\bbomega_{{\bm r}t} \times
 {\bf S}_{{\bm r}t}\right] -D\left[ \left( \Delta _{\bm r} {\bf S}_{{\bm r}t,\bot} \right) \times {\bf S}_{{\bm r}t} \right] , \nonumber  \\
\bbomega_{{\bm r}t}&\equiv \bomega_{Z{\bm r}} -\widetilde\omega_{\bm r} {\bf S}_{{\bm r}t,\bot}  +\Delta\bomega_t  ~ .   ~~~~~~~~~~~
\label{eq:spin-diffusion}
\end{align}
Here $\bomega_{Z{\bm r}}$ takes into account the non-uniform Lamb renormalization of $\omega_Z$ and the exchange contribution is transformed into $-\widetilde\omega_{\bm r}{\bf S}_{{\bm r}t,\bot}$, where ${\bf S}_{{\bm r}t,\bot}$ is the transverse part of spin orientation. The frequency $\widetilde\omega_{\bm r}$ is determined by the averaged exchange integral $\left\langle\sum_{kk'}\hat{J}^{(kk')}/\hbar\right\rangle$ with the non-zero and equal $xx$- and $yy$-components multiplied by number of donors in an effective volume of interaction. The diffusion coefficient, $D$, is estimated as $\widetilde\omega l_{ex}^2/2$ where $l_{ex}$ estimates a scale of exchange interaction. The diffusion contribution of Eq.~(\ref{eq:nu1}) is negligible for the case of large-scale ($\gg l_{ex}$) inhomogeneities of $\bomega_{Z{\bm r}}$ and $\widetilde\omega_{\bm r}$. Because the ratio $\hbar\nu_1 /J_{r}$ is negligible, we replace $\hat\gamma\cdot{\bf S}_{{\bm r}t}$ by $\nu_2{\bf S}_{{\bm r}t,\bot}$ with the transverse relaxation rate $\nu_2$.

Taking into account $\nu_2$ and neglecting diffusion if scale of disorder $>l_{ex}$, one obtains the nonlinear (with respect to ${\bf S}_{{\bm r}t}$ and $\Delta\bomega_t$) system for the transverse and longitudinal (${\bf S}_{{\bm r}t,\bot}$ and ${\bf e}_z S_{{\bm r}t,\|}$) parts of spin orientation
\begin{align}
\left(\frac{d}{dt}+\nu_2 \right){\bf S}_{{\bm r}t,\bot}&=\Omega_{{\bm r}t,\|}\left[ {\bf e}_z \times {\bf S}_{{\bm r}t,\bot}\right] - \Delta\bomega_t S_{{\bm r}t,\|} ~ , \nonumber  \\
\label{eq:spin-rotation}
\frac{d}{dt}S_{{\bm r}t,\|} &=\left(\Delta\bomega_t\cdot{\bf S}_{{\bm r}t,\bot }\right)  ~ . ~~~~~~~~~~~
\end{align}
Here $\Omega_{{\bm r}t,\|}=\omega_{Z{\bm r}}+\widetilde\omega_{\bm r}S_{{\bm r}t,\|}$ includes the Lamb shift and the exchange ($\propto S_{{\bm r}t,\|}$) renormalization, which can be parametrically dependent on $\bf x$ due to a large-scale disorder. Within the collisionless regime, $\nu_{2}t\ll 1$, the spin conservation takes place
${\bf S}_{{\bm r}t,\bot}^2 +S_{{\bm r}t,\|}^2=S_0^2$ with the ${\bm r}t$-independent initial orientation $S_0$. If $\Delta\bomega_t\to 0$ and $\nu_2\to 0$, Eq. (\ref{eq:spin-rotation}) describes free rotation of $S_{{\bm r}t,\bot}$ around $0Z$ with the frequency $\omega_{Z{\bm r}}+\widetilde\omega_{\bm r}S_0$. Characterization of PHC (exchange, relaxation, and disorder parameters) and manipulation of spins are possible under resonant mw pumping.

\subsection{Microwave response.} The absorbed power is given by $P_{{\bm r}t}=-\hbar\overline{\Delta\bomega_t\cdot d{\bf S}_{{\bm r}t,\bot }/dt}$, where $\overline{(\ldots )}$ means averaging over period $2\pi /\omega$.\cite{Schweiger:2001,Fanciulli:2009} Under weak circular pumping $\Delta\omega_{t,x} +i\Delta\omega_{t,y}= \omega_P\exp (i\omega t)$, the solution of linear Eq. (\ref{eq:spin-rotation}) gives the resonant peak $P_{\bm r}=\hbar\omega\omega_P^2 \nu_2 /\left[\nu_2^2 +(\delta\omega_{\bm r} -\widetilde\omega_{\bm r}/2 )^2 \right]$ where $\delta\omega_{\bm r}\equiv\omega -\omega_{Z{\bm r}}$ is the frequency detuning and $S_0=-1/2$ for the zero-temperature limit. The resonant line has linewidth determined by $\nu_2$ and by disordered contributions stem from $\delta\omega_{\bm r}$ and $\widetilde\omega_{\bm r}$. If line is narrow enough, these contributions can be verified from the shape of the differential absorption $dP/dH$ averaged over disorder, similarly to the measurements of GeSi dots.\cite{Zinovieva:2014} The derivative $dP/dH$ increases under the REI-conditions due to additional dependency of $\widetilde\omega$ on $\delta_Z$.

\begin{figure}[htbp]
\begin{center}
\includegraphics[scale=0.37]{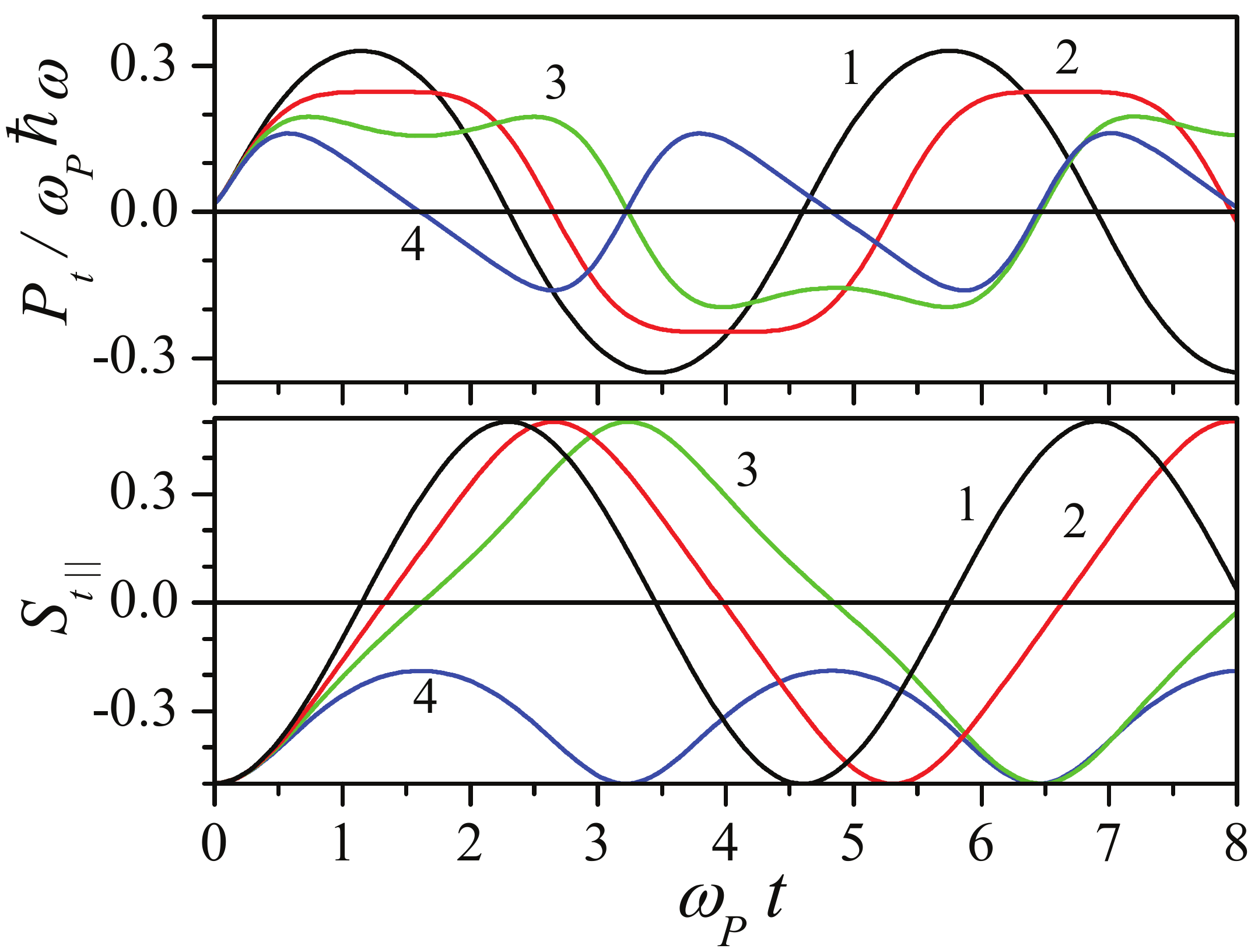}
\end{center}
\addvspace{-0.7 cm}
\caption{ {\bf Interplay between exchange and Rabi oscillations.} Absorbed power $P_t$ (in units $\omega_P\hbar\omega$) and the longitudinal spin orientation $S_{t,\|}$ versus dimensionless time, $\omega_Pt$, for $\widetilde\omega /\omega_P =$2 (1), 4 (2), 5 (3), 6 (4). }
\label{fig:Rabi}
\end{figure}

In the case of weak exchange, $\widetilde\omega\ll\max (\omega_P ,|\delta\omega |,\nu_{1,2} )$, the linear with respect to ${\bf S}_t\propto S_0$ system (\ref{eq:spin-rotation}) describes evolution of the resonant absorption $P_{{\bm r}t}$ and the spin orientation $S_{{\bm r}t,\|}$. Neglecting damping, at $\nu_2 t\ll 1$ and $\nu_2\ll \omega_P ,|\delta\omega_{\bm r} |$, and using the rotation wave approach, if $|\delta\omega |\ll \omega_Z$, one obtains oscillating responses
\begin{equation}
P_{{\bm r}t} =-S_0\hbar\omega\frac{\omega_P^2}{\omega_{{\bm r}R}}\sin\omega_{{\bm r}R} t ~ , ~~~   S_{{\bm r}t,\|}=S_0 \cos\omega_{{\bm r}R} t
\label{eq:Rabi}
\end{equation}
with the Rabi frequency $\omega_{{\bm r}R} =\sqrt{\omega_P^2 +\delta\omega_{\bm r}^2}$ and the $\pi /2$ phase shift between $P_{{\bm r}t}$ and $S_{{\bm r}t,\|}$. If the exchange interaction is essential ($\widetilde\omega\sim\omega_P ,|\delta\omega |$), shape and strength of temporal Rabi oscillations are sensitive to ratio $\widetilde\omega /\omega_P$. Within the rotating-wave approximation, we plot these responses in Fig.~\ref{fig:Rabi} for low temperatures, $S_0 =-1/2$ at resonant condition $\delta \omega =0$ (implicit solution for $S_{{\bm r}t,\|}$ can be written through the elliptic integrals).

The two-pulse spin echo scheme $(\pi /4$ - $\tau$ - $3\pi /4$ - $\tau\to$ echo) permits verification of exchange contribution under an essential decoherentization and long-range disorder.\cite{Schweiger:2001} Here two pulses of frequency $\omega$ and durations $t_{1,2}$ are correspondent to the pumping levels $\omega_{P1,2}$ by $\omega_{P1}t_1 =\pi /4$ and $\omega_{P2}t_2 =3\pi /4$ and the delay times $\tau\gg t_{1,2}$. In the rotating-wave frame, free evolution of ${\cal S}_{t,\bot}=\left\langle{S_{t,x}+iS_{t,y} }\right\rangle\exp (-i\omega t)$ after first and second pulses is $\propto\exp (\delta\omega_{1,2}t)$ with different frequencies $\delta\omega_{1,2}=\delta\omega_{\bm r}-\widetilde\omega S_{1,2\|}$ and $\left\langle\ldots\right\rangle$ stands for averaging over long-range disorder. If $\omega_{P1,2}\gg \widetilde\omega ,|\delta\omega_{\bm r}|, \nu_2$ and exchange is negligible according to Fig.~\ref{fig:Rabi}, the spin orientations after first and second pulses are $S_{1,\|}=S_0 /\sqrt{2}$ [c.f. Eq. (\ref{eq:Rabi})] and $S_{2,\|}=-S_0 (1+\cos\delta\omega_1\tau )/2$. For the case of Gaussian disorder with the averaged variations of Zeeman frequency $\sqrt{\left\langle{\delta\omega_{\bm r}^2}\right\rangle}=\delta\omega^*$, spin echo signal is
\begin{equation}
 {\cal S}_{t,\bot}= \overline{S}_0\exp \left\{  - \frac{\left[ {\delta \omega ^* (t-2\tau )} \right]^2 }{2} \right\}\Psi \left( {\phi _\tau  } \right) ,
\label{eq:spin-echo}
\end{equation}
where $\overline{S}_0=(1 + \sqrt 2 )/4$ stands for the echo amplitude at $t=2\tau$ if $\widetilde \omega\to 0$ and $\Psi (0)=1$. Function $\Psi \left( \phi_\tau \right)$ with $\phi_\tau =S_0\widetilde\omega\tau /\sqrt 2$ describes modulation of ${\cal S}_{t,\bot}$ caused by the exchange-induced difference in $\delta\omega_{1}$ and $\delta\omega_{2}$, which results in an interference oscillations of $\Psi \left( \phi_\tau \right)$. The disorder-induced exponent and the modulation $\Psi$ are multiplied because of additive contributions of these factors to the frequency $\Omega_{{\bm r}t,\|}$ in Eq.~(\ref{eq:spin-rotation}). Shape of modulation of ${\cal S}_{t=2\tau ,\bot}$ versus delay time $\tau$ is determined by ${\rm Re}\Psi$ and ${\rm Im}\Psi$ plotted in Fig.~\ref{fig:Rabi}. Here ${\rm Re}\Psi$ and ${\rm Im}\Psi$ are even and odd  functions of $\phi_\tau$ and $\phi_\tau >0$ corresponds to the spin inversion case, $S_0 >0$. Thus, verification of exchange contribution require variations of $\tau$ in $\sim 10$ $\mu$s scales.
\begin{figure}[thbp]
\begin{center}
\includegraphics[scale=0.37]{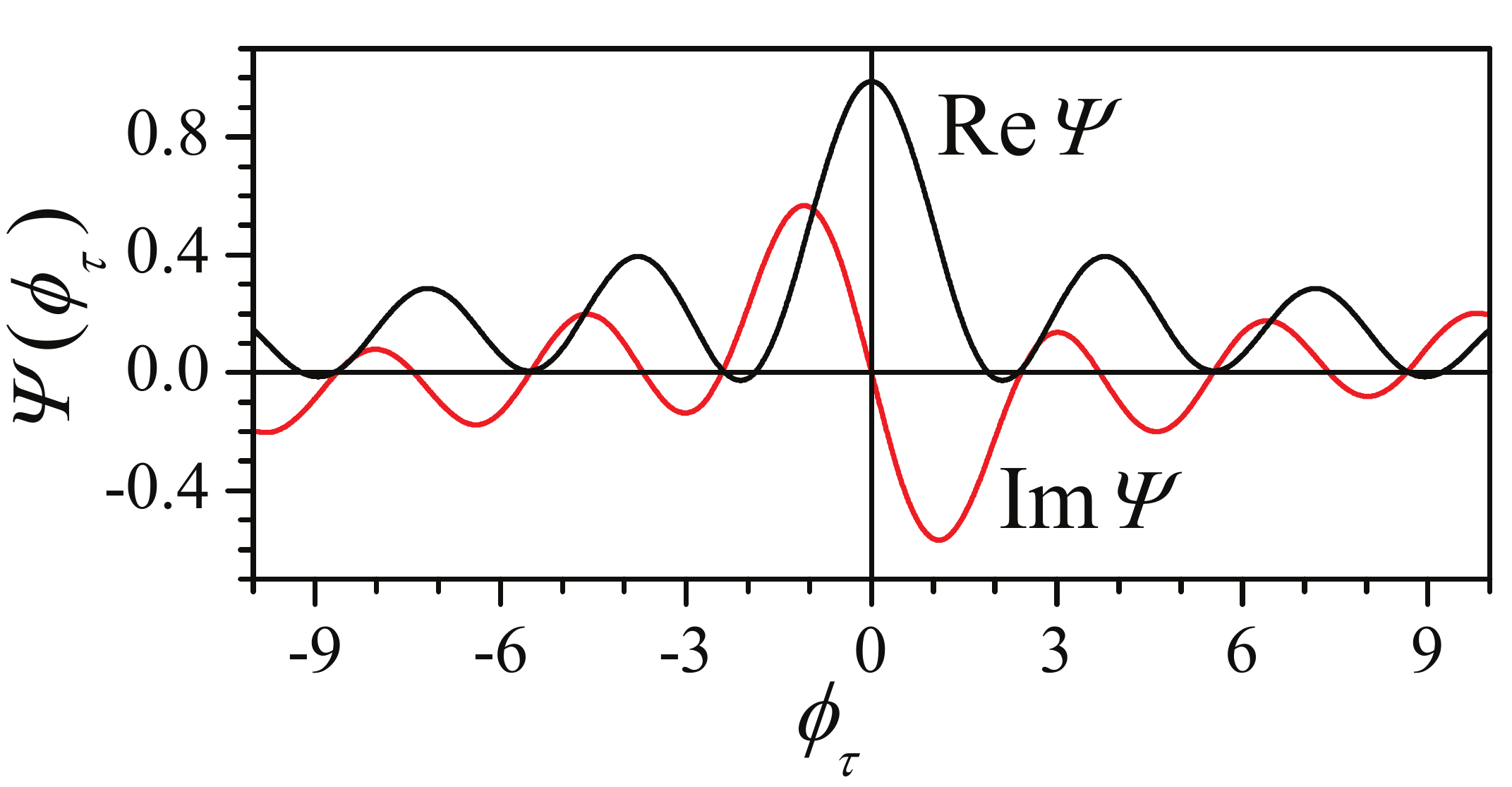}
\end{center}
\addvspace{-0.7 cm}
\caption{{\bf Oscillations of spin echo amplitude.} Real and imaginary parts of $\Psi$ versus delay time $\tau$ and $S_0$ (here $\phi_\tau =S_0\widetilde\omega\tau /\sqrt 2$). }
\label{fig:spin-echo}
\end{figure}
  
\subsection{Structure with quasi-gap.} 
\begin{figure}[thb]
\begin{center}
\includegraphics[scale=0.55]{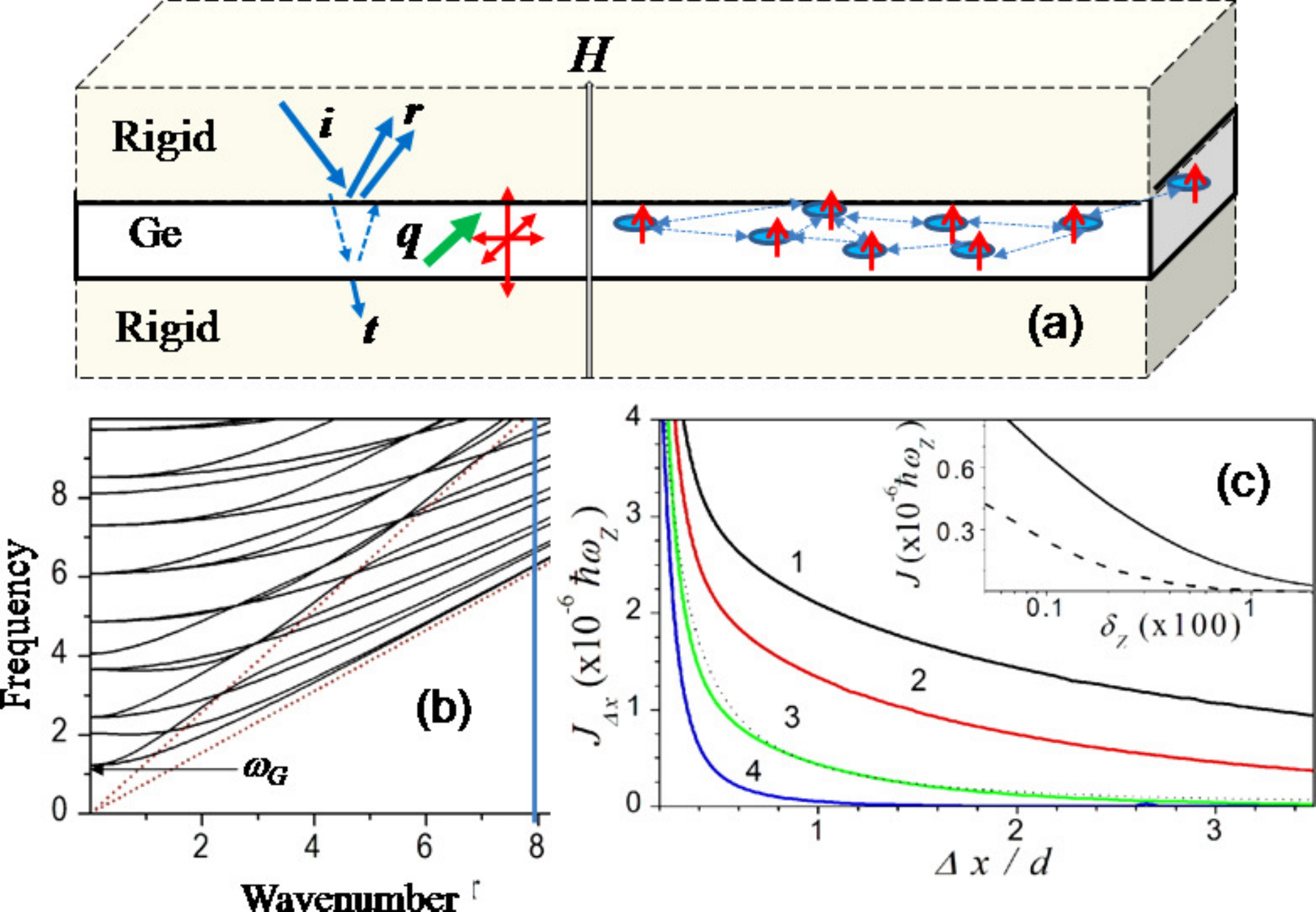}
\end{center}
\caption{{\bf Exchange  in sandwiched Ge structure.} ({\bf a}) r/Ge/r structure with interacted donor 
spins under magnetic field $\bf H$; incident ($i$-), reflected ($r$-), and transmitted ($t$-) waves and 
three waveguide modes (shear and coupled) with in-plane wave vector $\bf q$ are shown. ({\bf b}) 
Coupled eigenmodes of frequencies $\omega_{\nu q}$ for structure with Ge layer of width $d$ (cut-
off frequency $\omega_G\propto d^{-1}$); dashed lines correspond to the bulk dispersion laws for $l
$- and $t$-phonons. ({\bf c}) $J_{r}$ versus $\Delta x$ (in units $10^{-6}\hbar\omega_Z$  and
 $d$, respectively) for Ge layer sandwiched between rigid slabs under relative detunings $
 \delta_Z^2$: 10$^{-3}$ (1), $5\times 10^{-3}$(2), 0.05 (3), and 0.5 (4). Asymptotics $J_{r}
 \propto \Delta x^{-1.5}$ is shown by dotted curve and inset demonstrate $\ln$-dependency of 
 $J_{r}/\hbar\omega_Z$ on $\delta_Z$ for $\pi\Delta x/d\simeq$5 and 10 (solid and dashed 
 curves, respectively). }
\label{fig:sandwich}
\end{figure}

Finally, we describe the r/Ge/r-structure with Ge layer sandwiched between the rigid substrate and cover layers which can be diamond, BN, or H-SiC (see Fig.~\ref{fig:sandwich}a). Penetration of vibrations from rigid materials into Ge is weak because the reflection is effective due to the about 10 times differences\cite{Levinstein:1996} between modules of elasticity in rigid materials and soft Ge. Thus, the spin-phonon interaction with bulk modes is ineffective and there is a quasi-gap for waveguide modes in Ge layer up to cut-off frequency in GHz range ($\omega_G\approx c_t\pi /d$, see Fig.~\ref{fig:sandwich}b).
Calculations of the exchange integral~(\ref{eq:Jnm}) was performed for the r/Ge/r-structure neglecting penetration of waveguide modes into the rigid materials. In Fig.~\ref{fig:sandwich}c we show the ratio $J_{r}/\hbar\omega_Z$ versus inter-donor distance for different $\omega_Z /\omega_G$, which demonstrates $\propto\Delta x^{-3/2}$ asymptotic if $\Delta x/d\sim 1$. These results are in agreement with the estimates of $J_{r}$
given by Eq.~(\ref{eq:large-distance})
If $\pi\delta_Z\geq 10^{-(3\div 4)}$ (restriction due to disorder effect), the REI-regime is realized over a wide interval of $\Delta x$, between 10 nm and 0.01$\div$0.1 cm with a sharp suppression of the RIE-regime if $\delta_Z$ increases, see inset.
	
At low ($\ll \hbar\omega_Z$) temperatures, the longitudinal relaxation rate $\nu_1$ in the quasi-gap region $\omega_Z <\omega_G$ is determined by the spin-flip transitions via bulk modes, weakly propagated through r/Ge/r-structure. Based on the golden rule, we estimate the relative rate $\nu_1 /\omega_Z\propto K^2$  
and the result is:
\begin{equation}
\frac{\nu_1}{\omega_Z}\sim K^2\frac{\hbar\omega_Z}{
M
v_s^2
} ,  ~~M
= 
2\pi 
\varrho_M\left(\frac{
v_s}{\omega_Z} \right)^3  ~,
\label{eq:nu1-omegaZ}
\end{equation}
where the characteristic sound velocity $\widetilde{c}$ is combined from $c_{l,t}$ in Ge and rigid materials and $\nu_1\propto
v_s^{~5}$. For the above parameters, we got $\nu_1 /\omega_Z\leq 10^{-11}$ and $\nu_1^{-1}$ exceeds seconds. The ratio $J_{r}/\hbar\nu_1$ exceeds $10^3$ and $\nu_1$ is not affect on the spin-spin exchange in structures with quasi-gap. Thus, the r/Ge/r-structure with quasi-gap, which is more simple technologically than PHC, can be interesting for QIP applications.




\section{Discussion and Outlook} 
It is important to stress that these results are based on the estimates written through the ratios of $K^2\hbar\omega_Z$ to the characteristic energies $M v_s^2$, (see Eqs. 
(\ref{eq:nu-omega}) and (\ref{eq:nu1-omegaZ})), which is evident from the dimensional requirements. The evaluation of $\hat H_{SS}$ is restricted by donor concentrations $n_D\ll 10^{16}$ cm$^{-3}$, when the exchange due to tunneling overlap of donors is negligible. For lower concentrations, the mean-field approximation for exchange in Eqs.~(\ref{eq:Bloch})-(\ref{eq:spin-rotation}) is valid if $n_D^{-1/3}<$ radius of interaction. A region of intermediate concentrations, between the mean-field regime and a system of free donor spins (if $n_D\leq 10^{11}$ cm$^{-3}$) requires a special analysis.

Summarizing the results obtained, we have demonstrated that a controllable manipulation of in donor spin system placed into PHC is possible by the reasons:  
\begin{enumerate}[{\bf (i)}]
\item Strong suppression of relaxation-to-exchange ratio (in contrast to the bulk case\cite{Solenov:2007}) opens a way for fault-tolerant operations;  
\item Very sharp (at detunings $\delta_Z\leq 10^{-3}$ corrrespondent a weak variations of magnetic field) transformation from free spin system to large-scale REI-regime permits a remote control of qubits;
\item Effective control of spin conversion between ${\bf S}_{t,\bot}$ and $S_{t,\|}$ by microwave pulse takes place due to the interplay between exchange renormalization and Rabi oscillations;  
\item Formation of macroscopic spin patterns, with lateral sizes up to 0.1 mm, using micromagnets in order to control REI-regime, see \cite{Pioro:2008} and references therein;
\item Manipulation of single spin in ultra-pure Ge\cite{Haller:1981,Scappucci:2011} or spin clusters in small-size ($\geq 10~\mu$m) PHCs, which are governed by the nonlinear Bloch equations (\ref{eq:Bloch}) or (\ref{eq:spin-rotation}), employing ferromagnetic tip\cite{Bode:2003} and/or mw cavity,\cite{VanLoo:2013,Sigillito:2014} 
 without any electric circuits when noises are suppressed.
\end{enumerate}

In order to implement the structures suggested and to measure the peculiarities found, one have to meet several technological requirements for PHC structures: {\bf a)} suppression of spin decoherentization rate $\nu_2$ in comparison to typical values in Ge-based materials\cite{Claeys:2011,Vrijen:2000} {\bf b)} reduced stresses, dislocations, and interface disorder, {\bf c)} homogeneity of donor distribution in PHC or r/Ge/r-structures or controllable selective doping,\cite{Shinada:2005} far from imperfections at boundaries or interfaces, and {\bf d)} spatio-temporal stability of magnetic field and pumping characteristics, i.e. $\omega_Z$, $\omega$, and $\omega_P$, as well as frequency of gap edge allowed realization of REI-regime.

To conclude, we have demonstrated that a combination of phonon engineering provided gap in vibration spectra in PHC with unique spin and technological characteristics of Ge opens a way for quantum information applications. We believe that our paper will stimulate effort for preparation of structures suggested and for verification of qubit parameters. \\


\begin{thebibliography}{99}
\bibitem{Ladd:2010}
Ladd, T. D., Jelezko, F., Laflamme, R., Nakamura, Y., Monroe, C. \& O’Brien, J. L. Quantum computers. {\it Nature} {\bf 464}, 45-53 (2010).
\bibitem{Kloeffel:2013}
Kloeffel, C. \& Loss, D. Prospects for spin-based quantum computing in quantum dots. {\it Annu. Rev. Condens. Matter Phys.} {\bf 4}, 51-81 (2013).
\bibitem{Kane:1998}
Kane, B. E. A silicon-based nuclear spin quantum computer. {\it Nature} {\bf 393}, 133-137 (1998).
\bibitem{Dutt:2007}
Dutt, M. V. G., Childress, L., Jiang, L., Togan, E., Maze, J., Jelezko, F., Zibrov, A. S., Hemmer, P. R. \&  Lukin, M. D. Quantum Register Based on Individual Electronic and Nuclear Spin Qubits in Diamond. {\it Science} {\bf 316}, 1312-1316
(2007).
\bibitem{Smelyanskiy:2005}
Smelyanskiy, V.~N., Petukhov, A. G. \& Osipov, V.~V. Quantum computing on long-lived donor states of Li in Si. {\it Phys. Rev.} B {\bf 72}, 081304 (2005). 
\bibitem{Tyryshkin:2012}
Tyryshkin,   A. M., Tojo, S., Morton, J. J. L.,  Riemann, H., Abrosimov, N. V., Becker, P., Pohl, H.-J., Schenkel, T., Thewalt, M. L. W.,    Itoh, K. M. \& Lyon, S. A. Electron spin coherence exceeding seconds in high-purity silicon. {\it Nature Materials}  {\bf 11}, 143 (2012).
\bibitem{Claeys:2011}
Claeys, C. \& Simoen, E. {\it Germanium-Based Technologies: from Materials to Devices} (Elsevier, Amsterdam, 2011).
\bibitem{Liu:2000} 
Liu Z., Zhang X., Mao Y., Zhu Y. Y., Yang Z., Chan C. T. \& Sheng P., Locally resonant sonic materials, {\it Science}, {\bf 289}, 1734 (2000).
\bibitem{Yablonovich:1987} 
Yablonovitch E., Inhibited spontaneous emission in solid-state physics and electronics, {\it Phys. Rev. Lett.} {\bf 58}, 2059 (1987).
\bibitem{Yang:2004} 10. Yang S., Page J.H., Liu Z., Cowan M. L., Chan C. T. \& Sheng P., Focusing of sound in a 3D phononic crystal, {\it Phys. Rev. Lett.} {\bf 93}, 024301 (2004).
\bibitem{Jia:2010} 11. Jia G. \& Shi Z., A new seismic isolation system and its feasibility study, {\it Earthquake Engineering and Engineering Vibration} {\bf 9}, 75 (2010).
\bibitem{Maldovan:2006} Maldovan M.  \& Thomas E. L., Simultaneous localization of photons and phonons in two-dimensional periodic structures, {\it Appl. Phys. Lett.} {\bf 88}, 251907 (2006).
\bibitem{Alegre:2011} Alegre T. P. M., Safavi-Naeini A., Winger M. \& Painter O., Quasi-two-dimensional optomechanical crystals with a complete phononicbandgap, {\it Opt. Express} {\bf 19}, 5658 (2011).
\bibitem{Safavi:2011}
Safavi-Naeini, A. H., Mayer Alegre, T. P.,    Chan, J., Eichenfield, M., Winger, M.,  Lin, Q.,    Hill, J. T.,    Chang, D. E. \& Painter, O. Electromagnetically induced transparency and slow light with optomechanics. {\it Nature} {\bf 472}, 69 (2011).
\bibitem{Bode:2003}
Bode, M. Spin-polarized scanning tunnelling microscopy. {\it Rep. Prog. Phys.} {\bf 66}, 523 (2003).  
\bibitem{Haller:1981}
Haller, E.E., Hansen, W.L. \& Goulding, F.S. Physics of ultra-pure germanium {\it Adv. in Phys.} {\bf 30}, 93-138 (1981).
\bibitem{Shinada:2005}
Shinada, T., Okamoto, S., Kobayashi, T., \& Ohdomari, I., Enhancing semiconductor device performance using ordered dopant arrays. {\it Nature} {\bf 437}, 1128 (2005).
\bibitem{VanLoo:2013}
Van Loo, A. F., Fedorov, A., Lalumire, K., Sanders, B.C.,  Blais, A., \& Wallraff, A. Photon-Mediated Interactions Between Distant Artificial Atoms. {\it Science}, {\bf 342}, 1494 (2013).
\bibitem{Sigillito:2014}
Sigillito, A. J., Malissa, H., Tyryshkin, A. M., Riemann, H. , Abrosimov, N. V.,  Becker, P., Pohl, H.-J., Thewalt, M. L. W., Itoh, K. M., Morton, J. J. L., Houck, A. A.,  Schuster, D. I. \& Lyon, S. A. Fast, low-power manipulation of spin ensembles in superconducting
microresonators. {\it Appl. Phys. Lett.} {\bf 104}, 222407 (2014).
\bibitem{Schweiger:2001}
Schweiger A. \& Jeschke, G. {\it Principles of Pulse Electron Paramagnetic Resonance} (Oxford University Press, Oxford, 2001).
\bibitem{Roth:1960}
Roth, L. {\it g} Factor and Donor Spin-Lattice Relaxation for Electrons in Germanium and Silicon. {\it Phys. Rev.} {\bf 118} 1534-1540 (1960).
\bibitem{Hasegawa:1960}
Hasegawa, H., Spin-Lattice Relaxation of Shallow Donor States in Ge and Si through a Direct Phonon Process. {\it Phys. Rev.} {\bf 118}, 1523-1534 (1960).
\bibitem{Wilson:1964} Wilson, D.~K. Electron Spin Resonance Experiments on Shallow Donors in
Germanium. {\it Phys. Rev.} {\bf 134},  A265 (1964).
\bibitem{Barenco:1995}
Barenco A., Bennett C. H., Cleve R., DiVincenzo D. P., Margolus N., Shor P., Sleator T., Smolin J. A., and Weinfurter, H.~Elementary Gates for Quantum Computation. {\it Phys. Rev.} A {\bf 52} 3457 (1995).
\bibitem{Solenov:2007}
Solenov, D., Tolkunov, D. \& Privman, V. Exchange interaction, entanglement, and quantum noise due to a thermal bosonic field. {\it Phys. Rev.} B {\bf 75} 035134 (2007).
\bibitem{Redfield:1957}
Redfield, A.~G. On the Theory of Relaxation Processes. {\it IBM J. Res. Develop.} {\bf 1}, 19 (1957). 
\bibitem{Breuer:2007} Breuer, H. P. and Petruccione F., {\it The Theory of Open Quantum
Systems.} (Oxford University Press, New York, 2007).
\bibitem{Dyakonov:2008}
{\it Spin Physics in Semiconductors}, Ed. Dyakonov, M.I. (Springer, Berlin/Heidelberg, 2008).
\bibitem{Fanciulli:2009}
{\it Electron Spin Resonance and Related Phenomena in Low-Dimensional Structures}, Ed. Fanciulli, M. (Springer, Berlin 2009).
\bibitem{Zinovieva:2014}
Zinovieva, A. F., Stepina, N. P., Nikiforov, A. I., Nenashev, A. V.,  Dvurechenskii, A. V., Kulik, L. V., Carmo, M. C. \& Sobolev, N. A. Spin relaxation in inhomogeneous quantum dot arrays studied by electron spin resonance. {\it Phys. Rev.} B {\bf 89}, 045305 (2014).
%
\bibitem{Levinstein:1996} {\it Handbook Series on Semiconductor Parameters}, v. 1, Eds. Levinshtein, M., Rumyantsev, S., \& Shur, M. (World Scientific, 1996).
\bibitem{Pioro:2008}
Pioro-Ladrire, M., Obata, T., Tokura, Y., Shin, Y.-S.,  Kubo, T., Yoshida, K., Taniyama, T. \& Tarucha S. Electrically driven single-electron spin resonance in a slanting Zeeman field. {\it Nature Physics} {\bf 4}, 776 - 779 (2008)
\bibitem{Scappucci:2011}
Scappucci, G., Capellini, G., Johnston, B.,  Klesse, W. M. Miwa, J. A., \& Simmons M. Y.
A complete fabrication route for atomic-scale, donor-based devices in single-crystal germanium. {\it Nano Lett.} {\bf 11}, 2272 (2011).
\bibitem{Vrijen:2000}
Vrijen, R., Yablonovitch, E., Wang, K., Jiang, H. W., Balandin, A., Roychowdhury, V., Mor, T. \& DiVincenzo, D. Electron-spin-resonance transistors for quantum computing in silicon-germanium heterostructures. {\it Phys. Rev.} A, {\bf 62}, 012306 (2000). 
\bibitem{Stroscio:2001}
Stroscio, M. A., \& Dutta, M. {\it Phonons in Nanostructures.}  (Cambridge Univ. Press, 2001).
\end{thebibliography}
\end{document}